\documentclass[preprint2]{aastex}

\slugcomment{To appear in A.J. in abbreviated form.}

\shorttitle{NGC 6520}
\shortauthors{A. P. Odell}

\begin{document}


\title{A New Look at Open Cluster NGC 6520}


\author{Andrew P. Odell}
\affil{Dept of Physics and Astronomy, Northern Arizona University, Flagstaff AZ 86011}
\email{Andy.Odell@nau.edu}



\begin{abstract}
We use CCD and photoelectric photometry with Str{\"o}mgren filters along with medium resolution spectra to investigate NGC 6520, an open cluster very nearly in the direction of the galactic center.  We find an age of 60 Myr, a distance of 2 kpc, and an average reddening E(b-y) = 0.295, but which increases toward the south.  The average heliocentric radial velocity of the B stars is $-29$ km $s^{-1}$, while the velocity of the nearby Barnard 86 is about 0 (heliocentric, $-11$ km $s^{-1}$ compared to the LSR).  This velocity difference amounts to about 1.8 kpc since the cluster formed, implying that it is extremely doubtful NGC 6520 is related to Barnard 86.


\end{abstract}





\section{Introduction}

\begin{figure}[ht]
\epsscale{1.}
\plotone{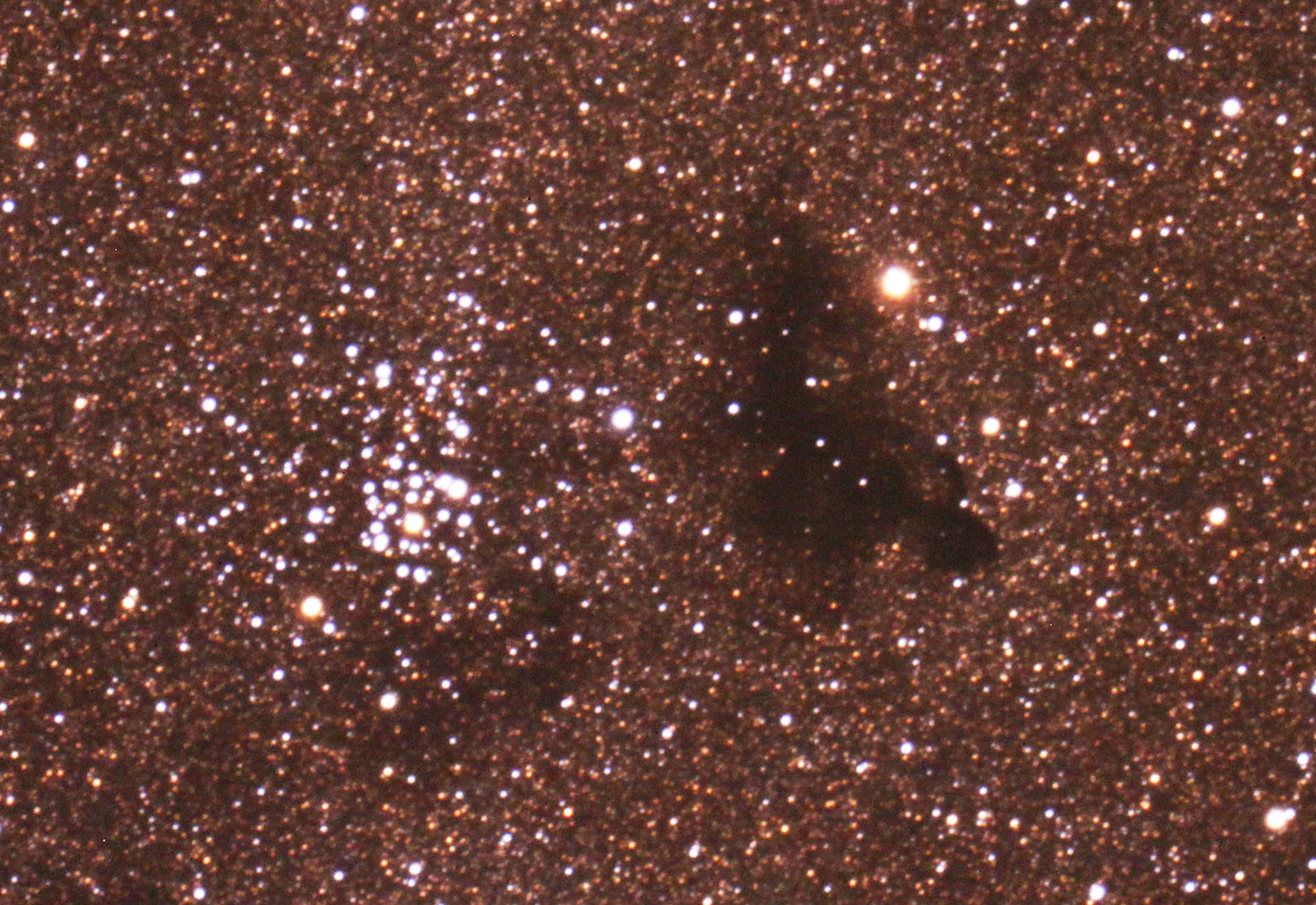}
\caption{An image taken by Klaus Brasch showing NGC 6520 and Barnard 86.  North is up, east to the left.  Note the patchy extinction visible to the south of the cluster.\label{fig1}} 
\end{figure}

The young open star cluster \objectname{NGC 6520} (C1800-279) has been neglected, mostly because it appears in a crowded field only four degrees from the Galactic center (see Fig.~\ref{fig1}).  The most recent study of the cluster, by Carraro et al. \cite{carraro}, derived an age of 150 $\pm$ 50 Myr (million years) and distance of 1.90 $\pm$ 0.1 kpc.    These authors noted that the nearby dark nebula Barnard 86 has a distance consistent with that of the cluster, and that the nebula overlaps the cluster.  They made the hypothesis that the two objects are related - that Barnard 86 is perhaps the cluster's birth cloud, which has become generally accepted.  This is puzzling, as it is thought that the mean lifetime of molecular clouds is $\approx$10 Myr \citep{blitz}.  

Hayford \cite{hayford}, in a study of Galactic rotation, published spectral types (from Trumpler) and radial velocities for three stars in \object{NGC 6520}.  Mermilliod et al. \cite{mermilliod} superseded that work with velocities for the three late type supergiants.

Zug \cite{zug}, in a followup study to Hayford, listed photographic magnitudes and color indices for 50 stars, along with spectral types (also from Trumpler) and color excesses for nine stars.  The existence of early B and possibly even an O star in the cluster argued for a very young age.  Houck \cite{houck} was analyzing a color-magnitude diagram (CMD) of the cluster, but the results were evidently never published.  Houck pointed out that there were three evolved supergiants in the direction of the cluster, which is rather unusual.

Svolopoulos \cite{svolo} produced a photographic CMD for NGC 6520, deriving a distance of 1.65 kpc and age of 800 Myr with a reddening estimate for E(B$-$V) of 0.27.  Lindoff \cite{lind} estimated a distance and age of 1.7 kpc and 54 Myr from relative numbers of stars on the upper main sequence, based on Zug's \cite{zug} data.  Santos Jr \& Bica \cite{santos} use integrated spectrophotmetry continuum levels and line strengths to estimate reddening of E(B$-$V)=0.56 and the low age of $\approx50$ Myr for \object{NGC 6520}.

The first believable photometry for NGC 6520 was done by Kjeldsen \& Frandsen \cite{kjeldsen} who used a CCD with UBV filters; they derive a distance of 1.6 kpc, a mean E(B$-$V) of 0.43, and they estimate the age to be 190 Myr; they note that the extinction could be variable over the cluster (which we confirm here).  Carraro et al. \cite{carraro} use this reddening value, as they did not obtain photometry in the U filter.  These disparate ages motivated the current study, which concludes that the cluster is younger than most previous estimates. 

\object{NGC 6520} was featured on the cover of the NOAO/NSO Newsletter September 2009 issue, with a comment that it '...probably formed from gas related to the nearby dust cloud.'  We find here that the cluster and the dust cloud were almost certainly not near each other when the cluster formed, even if they might be now.

\section{Observations}





\subsection{Photometry}  \label{phot}

NGC 6520 never rises above two airmass in Arizona and an inadequate set of standards was observed during the CCD observations discussed below. Therefore, calibration was done with photoelectric photometry (PE) from five nights in 1997 (UT June 29, 30 and July 1) and 1998 (UT July 1 and 2) using the White photometer with Str{\"o}mgren uvby filters on the Lowell Observatory 0.8m telescope.  Over 20 Str{\"o}mgren standard stars were used each night, and a total of 22 target stars were observed on at least two of the nights.  These targets were chosen on the basis of appearing single on preliminary CCD images, and with a range of color.

All five nights were extremely constant, photometric to within about 0.006 mag.  The calibration and reduction to standard magnitudes were done for each night using the PHOTCAL package of IRAF\footnote{IRAF is distributed by NOAO, which is operated by AURA under agreement with the National Science Foundation}, specifically the fitparams and evalfit tasks; color indices were used and a color term was included, and full documentation is available at the end of the longform of Table 1.  These magnitudes were used to further calibrate the CCD images as described next.

Multiple CCD images of NGC 6520 were obtained with the Steward Observatory 1.52m Kuiper telescope on UT April 3, 2000 using Str\"omgren uvby filters with the BigCCD 4K camera (binned 4x4 to give a field $12'$ square and pixel size $0.7''$).  The total exposure times were respectively 480, 480, 120, and 50 seconds in four or more images (to eliminate cosmic rays); twilight flats were used.  Eight Str{\"o}mgren standard stars were observed, which were not adequate for good calibration, so PE as described above was added to the calibration.

The CCD images were reduced in the standard way using IRAF, and the individual frames were combined.  Aperture photometry was done with IRAF task phot and the same aperture as for the PE data.  This allowed the PE cluster members to be used to derive the transformation to standard Str{\"o}mgren magnitudes (included at the end of the longform of Table 1) using the IRAF PHOTCAL tasks fitparams and invertfit.  

Final magnitudes were then extracted from each of the four combined images using DAOPHOT in IRAF and corrected to the standard system.  The standard colors thus formed are listed in Table~\ref{tbl-1} cols 4-7 (V (derived from y), $(b-y)$, $m_{1}$, and $c_{1}$).  IRAF formal errors are less than 0.04 mag for all colors of the B stars (and usually much smaller), and 0.02 mag in b and y for the A stars. Missing valuesof  $m_{1}$ and $c_{1}$ are due to large formal errors in the faint, red stars. Dereddening is described in section~\ref{analysis}.  

\subsection{Spectroscopy} \label{spec}

To better judge cluster membership, we obtained spectra of 30 stars near the cluster center on UT July 7, 2012, and a second set of spectra for nine of those on July 8 with the B\&C spectrograph on the Steward Observatory 2.3m Bok telescope. The second night was hampered by clouds, and the fact that good wavelength calibration was not available precludes radial velocities.  Spectra were also obtained for spectral standard stars.  The spectrograph configuration was such as to produce 1.35{\AA} \hspace{2pt} (two pixel) resolution, using the 832/mm grating in second order and a slit width of $1.5''$.  Wavelength coverage was about 3865\AA \hspace{2pt}to 4710\AA.  The spectra were reduced with IRAF and radial velocities were obtained with task fxcor. The IRAF task rvidlines was used to derive the velocities of $\tau$ Her (-1.0 km $s^{-1}$) and Zug 2 (-28.6 km $s^{-1}$), the templates for B stars and later types, respectively.

Spectral types and radial velocities are given in Table~\ref{tbl-1}, columns 11 and 12 respectively.  The fxcor uncertainty in velocity is about 10 km $s^{-1}$ for the B stars, and slightly smaller for later types.  Binarity is common among B stars; Chini et al. \cite{chini} find about 35\% binaries among B4-B6 and 20\% among B7-B9 stars.  With only one velocity for each case, we cannot identify individual binaries, but undetected companions will have an effect on velocities tabulated here.  In column 13, the letter N indicates the star may not be a cluster member, as discussed in section~\ref{members}.

Zug 2, 3, and 4 were included in the cluster RV program by Mermilliod et al. \cite{mermilliod}, who find -23.13, -24.12, and -23.41 km/sec, respectively. The values found here are -28.6, -32.5, and -23.5 km/sec, in reasonable agreement.  Our mean cluster velocity is also consistent with the average velocity of the cluster given by Hayford \cite{hayford} of $-26$ km $s^{-1}$, which was based on two spectra each of three stars, all having different values for the two spectra (i.e. likely binaries).

\section{Analysis} \label{analysis}
\subsection{Extinction}  \label{extinc}

De-reddening was done using the relation for unreddened B stars given in Crawford et al. \cite{crawford} $(b-y)_{0} = -0.116 + 0.097c_{1}$ by iterating in the following manner.  Based on the measured $c_{1}$, a preliminary $(b-y)_{0}$ was calculated with that equation.  That $(b-y)_{0}$ allowed for a preliminary color excess $E(b-y) = (b-y) - (b-y)_{0}$ to be found.  The extinction to $c_{1}$ is $E(c_{1}) = 0.2E(b-y)$ which allows a better $c_{1} = c_{1} - E(c_{1})$ to be calculated, and used in the first equation above to get an improved $(b-y)_{0}$; this is iterated until no further change occurs.  The de-reddened $(b-y)_{0}$ and $V_{0}$ are given in Table~\ref{tbl-1}, cols 8 and 9, and the color excess $E(b-y)$ is given in col 10.

\begin{figure}[ht]
\epsscale{1.0}
\plotone{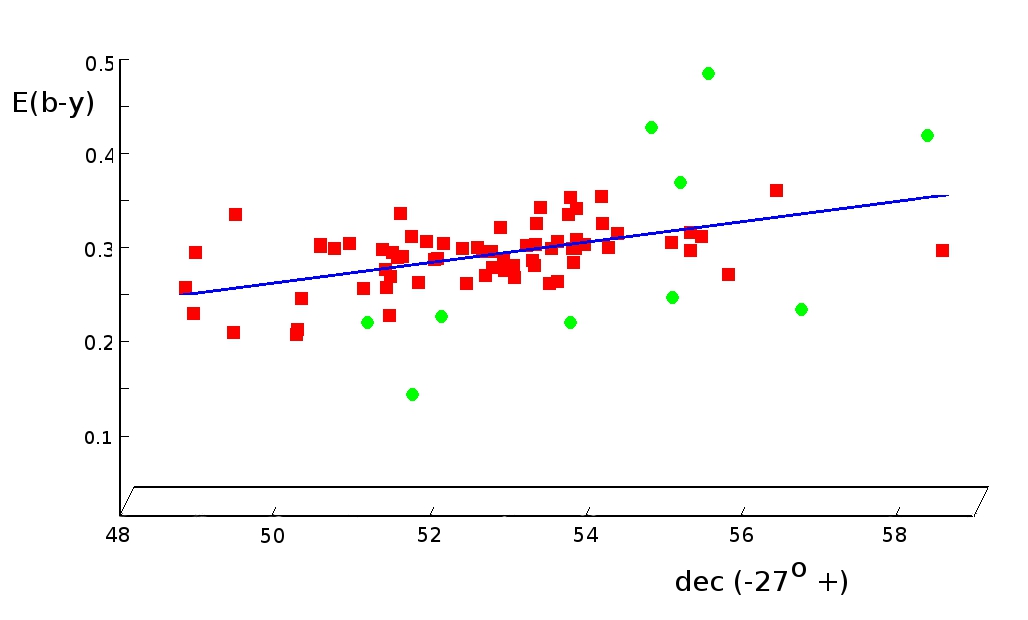}
\caption{Color Excess $E(b-y)$ in magnitudes for NGC6520 B stars (red squares) and probable non-members (green circles, not used in fit) as a function of declination in arcminutes south from $-27^{o}$. The (blue) line is the fit of a plane to the members.  The third axis of this 3-D plot is Right Ascension looking east into the page, from $18^{h}03^{m}00^{s}$ to $18^{h}03^{m}50^{s}$. The tilt of the RA axis has been chosen so that the reader is viewing the plane edge-on; this tilt is so slight that the RA values have been omitted from the third axis for lack of space in the figure.\label{fig2}}
\end{figure}

A least-squares plane fit of $E(b-y)$ vs RA and dec shows extinction decreasing eastward at a rate of $-0.71$ mmag per RA second, or $-2.5$ mmag per arcminute, and increasing southward by $10.9$ mmag per arcminute of dec.  This is shown in Fig.~\ref{fig2}, the derived $E(b-y)$ for B stars as a function of position on the sky (\textit{ie} a 3-D plot).  

Thus most of the extinction variation is in the north-south direction, even though Barnard 86 is located to the west.  This makes it unlikely that much or any of the extinction at the cluster is due to that cloud.  In an image of the cluster (see Fig.~\ref{fig1}), it is easy to see patchy clouds of dust in line with the southern border of the cluster, and thus the scatter in Fig.~\ref{fig2} is not unexpected, especially to the right.  The figure also shows some outlier stars where the extinction is anomalous and may be indicating these are not cluster members; see section~\ref{members}.

For stars later than B, the above method will not work to find extinction, so a least-squares fit of a plane was made to the extinction as a function of RA and dec for the B stars.  This fit was used to correct for extinction for later type stars in Table~\ref{tbl-1}, and for these stars, the $E(b-y)$ extinction in column 10 is marked with an asterisk (*).  Since the extinction is patchy, we expect this to give rise to substantially more scatter in the dereddened colors for the later type stars.

\subsection{Color Magnitude Diagram}  \label{cmd}

\begin{figure}[ht]
\plotone{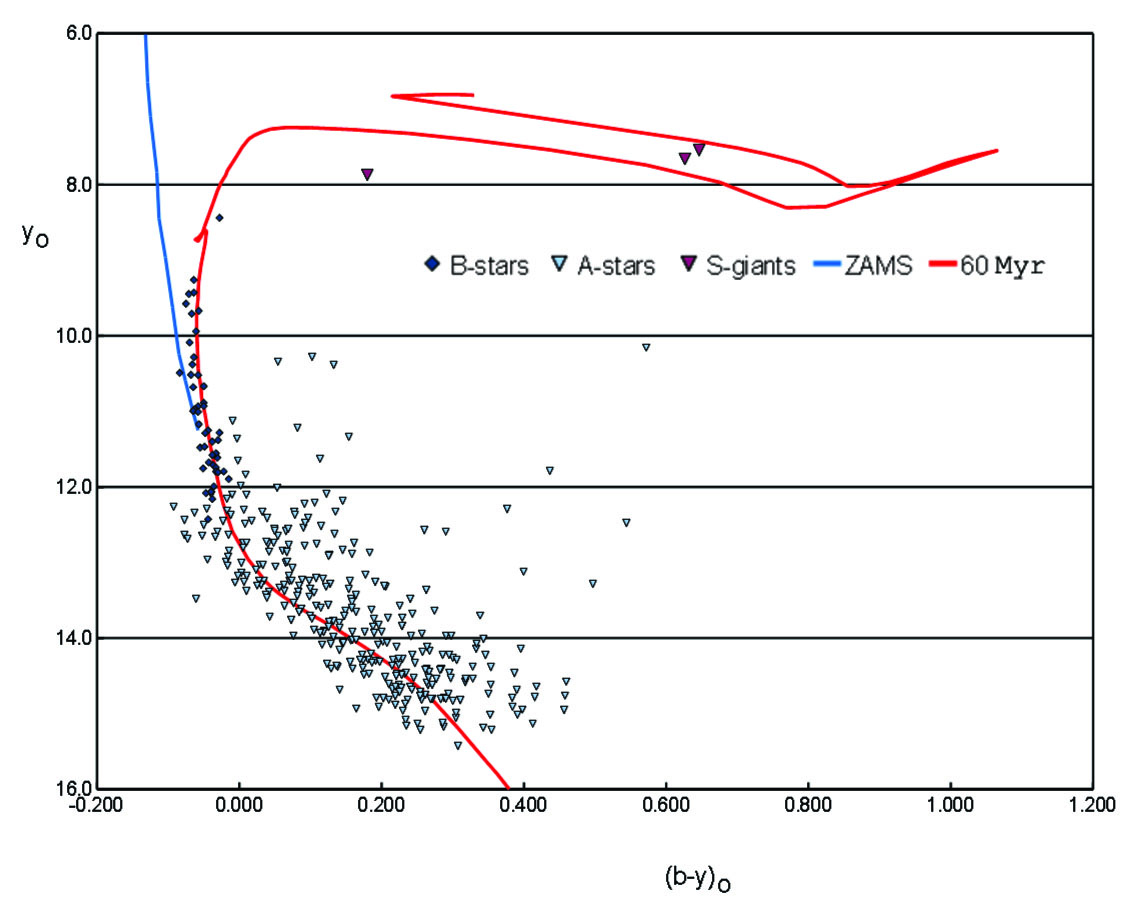}
\caption{The CMD for NGC 6520, showing $V_{0}$ vs $(b-y)_{0}$.  Included are the ZAMS and the 60 Myr isochrones, adjusted to a distance modulus of 11.5.\label{fig3}}
\end{figure}

A color-magnitude diagram (CMD) is an excellent diagnostic tool to determine age and distance for a star cluster.  Columns 8 and 9 ($(b-y)_{0}$ and $V_{0}$) of Table~\ref{tbl-1} are plotted in Fig.~\ref{fig3}, with different symbols for B stars dereddened individually, and the A stars and supergiants dereddened by a plane fit to the redenning.  As expected, the scatter for the B stars is much less than for the A stars.

Also plotted are two isochrones \footnote{(available at http://stev.oapd.inaf.it/cmd)}, lines of constant age, for stellar evolution models as described by Marigo et al. \cite{marigo}. The isochrones provide the uvby Str{\"o}mgren magnitudes for various mass stars at a chosen age and metallicity.  There are only a few adjustable parameters needed to fit the observed points, and so the age and distance that best represent the cluster can be determined.  The other parameter, metallicity, is chosen to be typical of recent star formation, Z=0.019.  Various ages were chosen, and the one that fit reasonably well is 60 Myr.  The B stars on the upper main sequence, for which reddening is well determined, are the prime ones to fit for age.  Further down the main sequence, in the region of the A stars, the isochrone can be adjusted up and down to fit the observed points by converting absolute magnitudes to the measured apparent magnitudes, adding a distance modulus, which is determined to be about 11.5, implying a distance of 2 kpc.  The two G supergiants, Zug 3 and 4, are very close to the blue loop in the isochrone.

The primary cause of uncertainty in the age and distance comes from the variable and patchy reddening.  However, the spectral types of the B stars indicate that the age of 60 Myr cannot be too far from correct; the earliest type is B4 V, indicating a mass of about 6$M_{\sun}$, which would have a main sequence lifetime of about this age.  It is unlikely that additional observations can improve the CMD analysis.

\begin{figure}[ht]
\epsscale{1.0}
\plotone{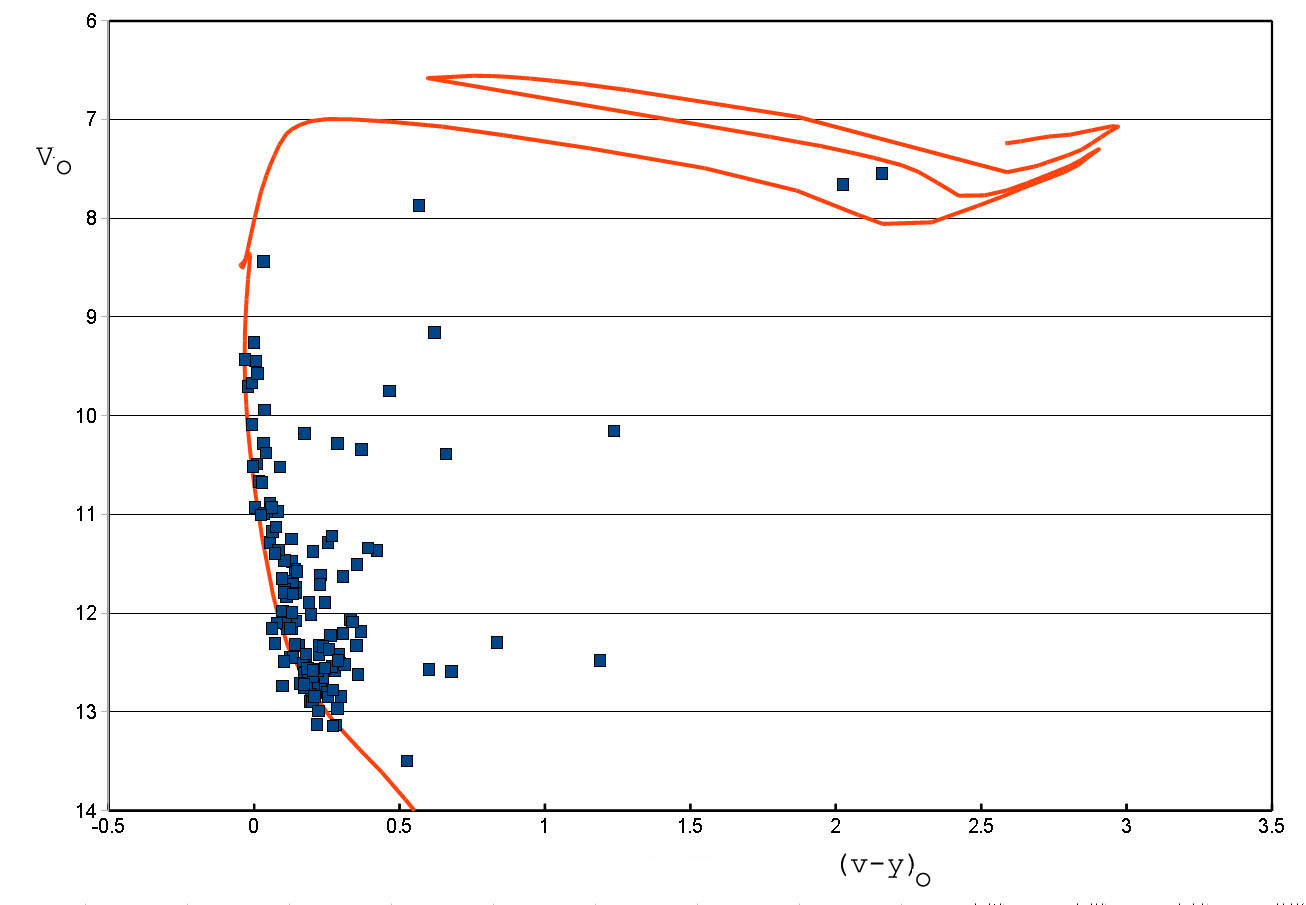}
\caption{The CMD for NGC 6520, showing $V_{0}$ vs $(v-y)_{0}$.  Included is the 60 Myr isochrone, adjusted to a distance modulus of 11.20.\label{fig4}}
\end{figure}

Fig.~\ref{fig4} is a CMD based on $(v-y)_{0}$ rather than $(b-y)_{0}$.  The range of this color is larger, and might be more discriminating of the age and distance to the cluster.  In this case, it can be seen that the age might be slightly greater than 60 Myr, and the best fit of the distance modulus is about 11.2, for a distance of 1.7 kpc.

We note here that the difference between our age and that of Carraro et al. \cite{carraro} arises from their exclusion of B stars farther than 30$''$ from the cluster center.  Our radial velocities show these stars to almost certainly be cluster members; they are the brightest, bluest stars in the cluster.  Further, dereddening individual B stars substantially decreases the scatter in our B star measurements.

\subsection{Cluster Membership}  \label{members}

There are several potential indicators in the data described here which could indicate cluster membership, such as radial velocity, extinction, strength of the CaII K line (for the B stars), position in the CMD, and/or distance from the cluster center (here taken to be $18^{h}03^{m}25^{s}$, $-27^{o}53'40''$).  The stars that are most interesting in this regard are as follows.

Zug 1 (=HD164621) is classified as a B9 III.  It is rather far from the cluster center, about $4'$ west.  Its extinction is the lowest of any B star in the cluster, $E(b-y) \approx 0.14$, where the average is about 0.27 mags at its dec.  The CaII line has an equivalent width of about 0.6\AA, whereas the typical B star in the cluster is more like 0.3\AA.  The radial velocity is about $-25$ km $s^{-1}$, similar to the average of the other B stars in the cluster of $-29$ km $s^{-1}$.  This being the brightest star in the cluster, it would be valuable as an age indicator, but not essential; it lies about a half-magnitude below the 60 Myr isochrone.  Taken together, all this indicates the star may not be a cluster member.

Zug 2 (=HD164654, Hayford 1) is an F4 or F5 Ib star located about an arc minute from the cluster center.  The extinction and Ca II line cannot indicate anything about cluster membership.  Its radial velocity is $-28.6$ km $s^{-1}$, near the cluster average.  This star's position in the CMD is also about a half magnitude below the first crossing to the red with a hydrogen burning shell.  This crossing is quite fast, so it would be unlikely to catch a star in this phase, but not impossible.

Zug 3 (=HD164684) and Zug 4 (=CD$-27$ 12315) are both G8 I spectral type, the former about two arc minutes southeast of the cluster center, the latter within a few arc seconds.  Their radial velocities are $-44$ and $-36$ km $s^{-1}$, both close to the average.  They both lie on the blue loop of the CMD, indicating they are in the core helium burning phase.

There are several other, less important stars that are probably not cluster members, based on radial velocity and extinction; they are indicated in column 10 of Table~\ref{tbl-1}.

\section{Conclusions}

We find NGC 6520 to be quite a bit younger than most previous estimates, i.e. about $60\pm10$ Myr, consistent with stars of spectral type B4 and B5 in the cluster \citep{negue}.  Its distance of about $2\pm0.2$ kpc (putting it on the outer edge of the Scutum-Centaurus spiral arm \citep{church}) has been reasonably estimated, but the distance to Barnard 86 has never been measured.  Under the assumption that Barnard 86 is at the same distance as the cluster, Carraro et al. \cite{carraro} found its properties to be reasonable, but this is hardly proof.  

They also measured a radial velocity of +11 km $s^{-1}$ compared to the local standard of rest, which is consistent with that estimated by Clemens \& Barvainis \cite{clemens}.  This translates into roughly 0.0 km $s^{-1}$ heliocentric velocity.  Table~\ref{tbl-1} col. 12 lists the heliocentric radial velocities of stars for which we obtained spectra; the typical velocity is about $-30$ km $s^{-1}$.  

The main conclusion here is that the cluster and cloud have different heliocentric radial velocities by about 30 km $s^{-1}$.  This, in the 60 Myr the cluster has been in existence, amounts to about 1.8 kpc relative motion, and would be greater if the cluster were older.  So, if the two objects are at the same distance today, they were not at the time the cluster formed, and thus almost certainly have nothing to do with each other.  Other problems with this idea include no known mechanism for ejecting an entire cluster from its birth cloud at that speed, and an ejection radially outward through the galaxy, but with essentially no tangential velocity.  We conclude \object{NGC 6520} was not formed in \object{Barnard 86}.  

It is possible that VLBI radio observations could establish a heliocentric parallax for Barnard 86, which would be interesting to determine its location and pin down some of its characteristics.

\acknowledgments
I thank Steward and Lowell Observatories for allocating telescope time for this project.  I appreciate many useful discussions with Brian Skiff, and correspondence with Daniel Majaess and Giovanni Carraro, as well as many useful comments from an anonymous referee.  This work made use of the SIMBAD database, operated at CDS, Strasbourg, France.

I would like to dedicate this paper to the memory of Dr. Theodore E. Houck, who taught me to be careful in everything I do.

{\it Facilities:} \facility{LO:0.8m (White photometer)}, \facility{SO:Kuiper (BigCCD)}, \facility{Bok (B\&C Spectrograph)}

\clearpage

\begin{figure}[ht]
\epsscale{2.4}
\plotone{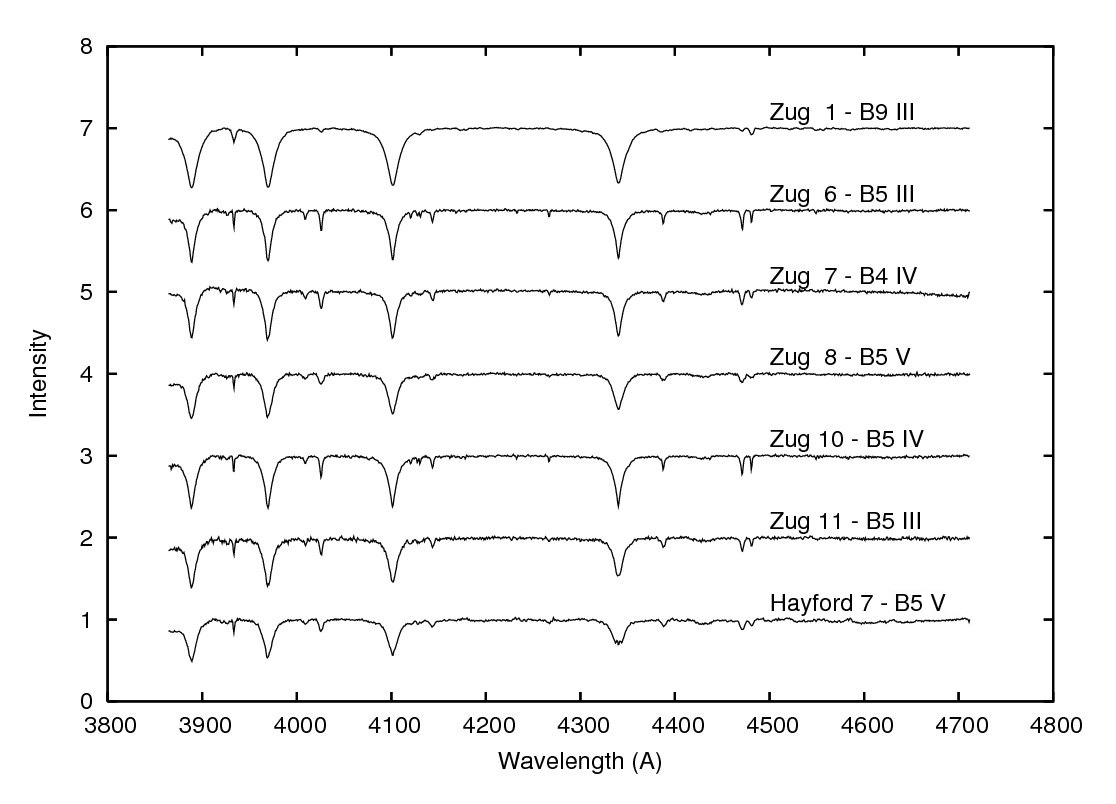}
\caption{Spectra of seven of the earliest and most interesting B stars in NGC6520.\label{fig5}} 
\end{figure}

\clearpage

\begin{figure}[ht]
\epsscale{2.4}
\plotone{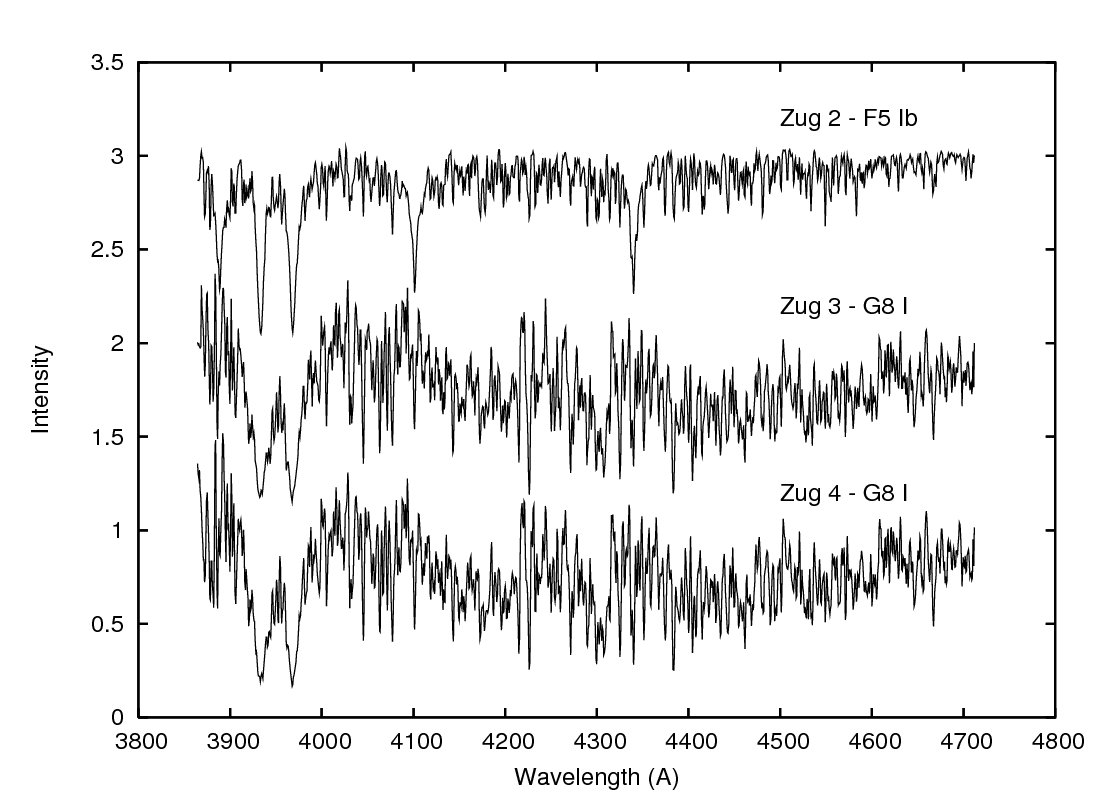}
\caption{Spectra of the three late-type supergiants in NGC6520.\label{fig6}} 
\end{figure}

\clearpage



\appendix


. Make sure there is at least one \tablenotemark
\tablecomments{Table \ref{tbl-1} is published here in its entirety.  In cases where the IRAF formal errors are large for m1 and c1, values have been dropped.}
\tablenotetext{a}{The * indicates extinction found from fit to B star extinctions.  See section~\ref{extinc}.}
\tablenotetext{b}{N means likely not a cluster member; see section~\ref{members}.}
\end{deluxetable}

\clearpage

\appendix

\section{Photometry Calibration Details}

In order to document the data reduction process which produced the photometry in this table, the following material is included.  Table~\ref{tbl-2} lists the Str\"omgren standard stars used for the PE photometry and their properties. Table~\ref{tbl-3} lists the Str\"omgren magnitudes and colors derived from the PE photometry for stars in the area of the cluster. 

\begin{deluxetable}{clllllllll}
\tabletypesize{\scriptsize}
\tablecaption{Standard Stars used for Photoelectric Photometry of NGC6520 \label{tbl-2}}
\tablewidth{0pt}
\tablehead{
\colhead{Star} & \colhead{V} & \colhead{$\Delta$ V } & \colhead{$(b-y)$} & \colhead{$\Delta (b-y)$} & \colhead{m1} & \colhead{$\Delta$ m1} & \colhead{c1} & \colhead{$\Delta$ c1} & \colhead{night\tablenotemark{a}} }
\startdata
\noalign{\vskip 0.5mm}
  BD043508 & 9.326 & 0.02  &  1.178 & 0.02  & INDEF &  INDEF & INDEF & INDEF & 123$-$$-$   \\
  HD110184 & 8.297 & 0.01  &  0.818 & 0.01  & 0.163 &  0.01  & 0.707 & 0.01  & 123$-$$-$   \\
  HD119537 & 6.518 & 0.02  &  0.029 & 0.02  & 0.172 &  0.02  & 0.983 & 0.02  & $-$$-$$-$45   \\
  HD120086 & 7.872 & 0.02  & $-0.083$ & 0.02  & 0.098 &  0.02  & 0.15  & 0.02  & $-$$-$$-$45   \\
  HD123825 & 7.254 & 0.02  &  0.984 & 0.02  & 0.792 &  0.02  & 0.395 & 0.02  & 123$-$$-$   \\
  HD126273 & 7.190 & 0.02  &  1.063 & 0.02  & 0.639 &  0.02  & 0.577 & 0.02  & 123$-$$-$   \\
  HD129956 & 5.685 & 0.02  &  0.005 & 0.02  & 0.120 &  0.02  & 1.023 & 0.02  & $-$$-$$-$4$-$   \\
  HD131597 & 8.429 & 0.01  &  0.473 & 0.01  & 0.206 &  0.01  & 0.303 & 0.01  & 123$-$$-$   \\
  HD136831 & 6.291 & 0.02  &  0.008 & 0.02  & 0.136 &  0.02  & 1.102 & 0.02  & $-$$-$$-$4$-$   \\
  HD140850 & 8.816 & 0.02  &  1.102 & 0.02  & INDEF &  INDEF & INDEF & INDEF & 123$-$$-$   \\
  HD145774 & 7.484 & 0.02  & $-0.021$ & 0.02  & 0.062 &  0.02  & 0.076 & 0.02  & $-$$-$$-$4$-$   \\
  HD154345 & 6.771 & 0.02  &  0.449 & 0.02  & 0.270 &  0.02  & 0.286 & 0.02  & 123$-$$-$   \\
  HD160233 & 9.095 & 0.02  &  0.025 & 0.02  & 0.032 &  0.02  & 0.071 & 0.02  & $-$$-$$-$45   \\
  HD160314 & 7.730 & 0.02  &  0.273 & 0.02  & 0.161 &  0.02  & 0.638 & 0.02  & 123$-$$-$   \\
  HD160315 & 6.260 & 0.02  &  0.642 & 0.02  & 0.392 &  0.02  & 0.452 & 0.02  & 123$-$$-$   \\
  HD161573 & 6.847 & 0.02  &  0.060 & 0.02  & 0.052 &  0.02  & 0.346 & 0.02  & $-$$-$$-$45   \\
  HD161817 & 6.982 & 0.01  &  0.137 & 0.01  & 0.113 &  0.01  & 1.206 & 0.01  & 123$-$$-$   \\
  HD162596 & 6.342 & 0.02  &  0.717 & 0.02  & 0.376 &  0.02  & 0.429 & 0.02  & 123$-$$-$   \\
  HD165401 & 6.801 & 0.01  &  0.393 & 0.01  & 0.166 &  0.01  & 0.288 & 0.01  & 123$-$$-$   \\
  HD165462 & 6.346 & 0.02  &  0.700 & 0.02  & 0.279 &  0.02  & 0.552 & 0.02  & 123$-$$-$   \\
  HD169578 & 6.730 & 0.02  &  0.049 & 0.02  & 0.073 &  0.02  & 0.860 & 0.02  & 123$-$$-$   \\
  HD172365 & 6.369 & 0.02  &  0.510 & 0.02  & 0.226 &  0.02  & 0.701 & 0.02  & 123$-$$-$   \\
  HD172829 & 8.460 & 0.02  &  1.389 & 0.02  & INDEF &  INDEF & INDEF & INDEF & 123$-$$-$   \\
  HD174240 & 6.235 & 0.02  &  0.037 & 0.02  & 0.131 &  0.02  & 1.112 & 0.02  & 123$-$$-$   \\
  HD176582 & 6.420 & 0.02  & $-0.068$ & 0.02  & 0.094 &  0.02  & 0.260 & 0.02  & 1234$-$   \\
  HD190299 & 5.670 & 0.02  &  0.827 & 0.02  & 0.537 &  0.02  & 0.470 & 0.02  & 123$-$$-$   \\
  HD198820 & 6.427 & 0.01  & $-0.053$ & 0.01  & 0.097 &  0.01  & 0.364 & 0.01  & $-$2345   \\
  HD199280 & 6.566 & 0.02  & $-0.031$ & 0.02  & 0.109 &  0.02  & 0.777 & 0.02  & $-$$-$$-$45   \\
  HD200340 & 6.498 & 0.02  & $-0.026$ & 0.02  & 0.082 &  0.02  & 0.559 & 0.02  & $-$$-$$-$$-$5   \\
  HD208527 & 6.389 & 0.02  &  1.091 & 0.02  & 0.743 &  0.02  & 0.426 & 0.02  & 123$-$$-$   \\
  HD210460 & 6.182 & 0.005 &  0.446 & 0.005 & 0.212 &  0.005 & 0.326 & 0.005 & 123$-$$-$   \\
\tableline

\enddata
\tablecomments{The standard values were taken from the Astronomical Almanac 2003 published by the US Naval Observatory, Appendix  ubvy and H-beta Standard Stars.}
\tablenotetext{a}{On some nights repeat observations were made.}
\end{deluxetable}

The transformation equations to standard magnitudes and colors for the PE photometry (values from 6/29/1997; other nights are similar): \\
\scriptsize 
y =	mag1 + (0.6204$\pm$0.0098) + (0.0327$\pm$0.0078)(mag2-mag1) - (0.1382$\pm$0.0072)X1 \hspace{120pt} stdev = 0.0058 \\
(b-y) =	(mag2-mag1) + (0.0321$\pm$0.0101) + (0.0778$\pm$0.0082)(mag2-mag1) - (0.0596$\pm$0.0070)X2 \hspace{80pt} stdev = 0.0050 \\
m1 = (mag3-mag2)-(mag2-mag1) + (0.1243$\pm$0.0108) - (0.1186$\pm$0.0091)(mag2-mag1) - (0.0613$\pm$0.0076)X3 \hspace{42pt} stdev =	0.0123 \\
c1 = (mag4-mag3)-(mag3-mag2) + (0.9481$\pm$0.0259) + (0.2222$\pm$0.0186)(mag2-mag1) - (0.1364$\pm$0.0177)X4 \hspace{40pt} stdev =	0.0313 \\
\normalsize

The transformation equations to standard magnitudes for the CCD photometry are: \\
\scriptsize 
mag1 = y + (4.3396 $\pm$ 0.0294) - (0.0070 $\pm$ 0.0634)by	\hspace{226pt} stdev = 0.0965 \\
mag2 = (by+y) + (4.4062 $\pm$ 0.0256) + (0.2001 $\pm$ 0.1094)by	\hspace{202pt} stdev = 0.0661 \\
mag3 = im1 + 2*by + y + (5.1438 $\pm$ 0.0305) + (0.1763 $\pm$ 0.1418)by	\hspace{171pt} stdev = 0.0623 \\
mag4 = c1 + 2*m1 + 2*by + y + (5.7482 $\pm$ 0.2119) + (0.2221 $\pm$ 0.2119)by \hspace{146pt} stdev = 0.0976 \\
\normalsize

\begin{deluxetable}{rrrrrrrrr}
\tabletypesize{\scriptsize}
\tablecaption{Calibrated Cluster Stars from Photoelectric Photometry of NGC6520 \label{tbl-3}}
\tablewidth{0pt}
\tablehead{
\colhead{Zug No} & \colhead{V} & \colhead{$\Delta$V\tablenotemark{a}} & \colhead{$(b-y)$} & \colhead{$\Delta(b-y)$} & \colhead{m1} & \colhead{$\Delta$m1} & \colhead{c1} & \colhead{$\Delta$c1}
 }

\startdata
\noalign{\vskip 0.5mm}
 8      &   10.904 & 0.004 &  0.233 & 0.002 &  0.073 & 0.021 &  0.487 & 0.028  \\
 -\tablenotemark{b} &   10.538 & 0.016 &  0.701 & 0.018 &  0.470 & 0.073 &  0.323 & 0.118  \\
 6      &   10.601 & 0.012 &  0.229 & 0.005 &  0.026 & 0.015 &  0.504 & 0.024  \\
 1      &    9.064 & 0.014 &  0.049 & 0.005 &  0.161 & 0.035 &  1.012 & 0.020  \\
 3      &    8.883 & 0.006 &  1.144 & 0.008 &  0.712 & 0.028 &  0.187 & 0.040  \\
 4      &    8.908 & 0.015 &  1.052 & 0.023 &  0.550 & 0.039 &  0.149 & 0.020  \\
 2      &    9.020 & 0.005 &  0.503 & 0.021 &  0.111 & 0.030 &  1.167 & 0.022  \\
10      &   11.155 & 0.040 &  0.214 & 0.026 &  0.065 & 0.040 &  0.470 & 0.004  \\
16      &   11.311 & 0.068 &  0.528 & 0.018 &  0.204 & 0.045 &  0.444 & 0.022  \\
17      &   11.752 & 0.020 &  0.290 & 0.020 &  0.022 & 0.065 &  0.654 & 0.067  \\
11      &   11.306 & 0.023 &  0.303 & 0.019 &  0.017 & 0.037 &  0.586 & 0.017  \\
34      &   12.396 & 0.083 &  0.366 & 0.045 &  0.006 & 0.083 &  0.694 & 0.053  \\
 9      &   10.790 & 0.008 &  0.234 & 0.058 & -0.033 & 0.103 &  0.754 & 0.067  \\
 5      &   10.210 & 0.006 &  0.194 & 0.008 & -0.030 & 0.002 &  0.653 & 0.023  \\
14      &   11.587 & 0.011 &  0.326 & 0.021 &  0.127 & 0.018 &  0.564 & 0.016  \\
15      &   11.772 & 0.030 &  0.140 & 0.044 &  0.083 & 0.015 &  0.507 & 0.025  \\
22      &   12.223 & 0.062 &  0.120 & 0.018 &  0.097 & 0.016 &  0.943 & 0.011  \\
12      &   11.404 & 0.011 &  0.104 & 0.001 &  0.077 & 0.006 &  0.436 & 0.039  \\
29      &   12.478 & 0.011 &  0.071 & 0.015 &  0.134 & 0.025 &  0.809 & 0.030  \\
24      &   12.187 & 0.022 &  0.286 & 0.044 &  0.043 & 0.057 &  0.568 & 0.022  \\
27      &   12.531 & 0.012 &  0.176 & 0.016 &  0.033 & 0.028 &  0.778 & 0.034  \\
28      &   12.488 & 0.010 &  0.312 & 0.015 &  0.064 & 0.026 &  0.995 & 0.037  \\

\tableline

\enddata

\tablenotetext{a}{The $\Delta$V etc values are the standard deviation of a single observation from different nights.}
\tablenotetext{b}{The Foreground K giant; RA 18 03 25.5 dec -27 55 58.}
\end{deluxetable}



\end{document}